\documentclass[final,1pt]{elsarticle}

\usepackage{amssymb}
\usepackage[utf8]{inputenc}
\usepackage{color}
\usepackage{amsmath}
\usepackage{amssymb} 
\usepackage{url}
\usepackage[margin=2.5cm]{geometry}
\usepackage{graphicx}
\journal{Computers in Biology and Medicine}

\begin{document}

\newcommand{\NW}[1]{{\color{black}{#1}}}
\newcommand{\RR}[1]{{\color{black}{#1}}}
\newcommand{\T}[1]{{\color{black}{#1}}}
\newcommand{\TT}[1]{{\color{black}{#1}}}

\begin{frontmatter}

\title{An Incremental Learning Approach to Automatically Recognize Pulmonary Diseases from the Multi-vendor Chest Radiographs}

\author[a,c]{Mehreen Sirshar}
\author[a,b,c,d]{Taimur Hassan}
\author[a]{Muhammad Usman Akram}
\author[a]{Shoab Ahmed Khan}

\affiliation[a]{organization={Department of Computer and Software Engineering},%Department and Organization
            addressline={National University of Sciences and Technology}, 
            city={Islamabad},
            postcode={44000}, 
            country={Pakistan}}

\affiliation[b]{organization={Center of Cyber-Physical Systems, Department of Electrical Engineering and Computer Sciences},%Department and Organization
            addressline={Khalifa University}, 
            city={Abu Dhabi},
            postcode={127788}, 
            country={United Arab Emirates}
            }

\affiliation[c]{Mehreen Sirshar and Taimur Hassan have contributed equally in the manuscript.}

\affiliation[d]{Corresponding Author, Email: taimur.hassan@ku.ac.ae}

\begin{abstract}
The human respiratory network is a vital system that provides oxygen supply and nourishment to the whole body. Pulmonary diseases can cause severe respiratory problems, leading to sudden death if not treated timely. Many researchers have utilized deep learning systems (in both transfer learning and fine-tuning modes) to diagnose pulmonary disorders using chest X-rays (CXRs). However, such systems require exhaustive training efforts on large-scale (and well-annotated) data to \TT{effectively} diagnose chest abnormalities (at the inference stage). Furthermore, procuring such large-scale data (in a clinical setting) is often infeasible and impractical, especially for rare diseases. With the recent advances in incremental learning, researchers have periodically tuned deep neural networks to learn different classification tasks with few training examples. \TT{Although, such systems can resist catastrophic forgetting, they treat the} knowledge representations (which the network learns periodically) independently of each other, and this limits their classification performance. Also, to the best of our knowledge, there is no incremental learning-driven image diagnostic framework (to date) that is specifically designed to screen pulmonary disorders from the CXRs. To address this, we present a novel framework that can learn to screen different chest abnormalities incrementally (via few-shot training). In addition to this, the proposed framework is penalized through an incremental learning loss function that infers Bayesian theory to recognize structural and semantic inter-dependencies between incrementally learned knowledge representations to diagnose the pulmonary diseases effectively (at the inference stage), regardless of the scanner specifications. We tested the proposed framework on five public CXR datasets containing different chest abnormalities, where it achieved an accuracy of 0.8405 and the F1 score of 0.8303, outperforming various state-of-the-art incremental learning schemes. It also achieved a highly competitive performance compared to the conventional fine-tuning (transfer learning) approaches while significantly reducing the training and computational requirements.

\end{abstract}


\end{frontmatter}

\section{Introduction}
\begin{figure}[t]
\centering
\includegraphics[scale=0.4]{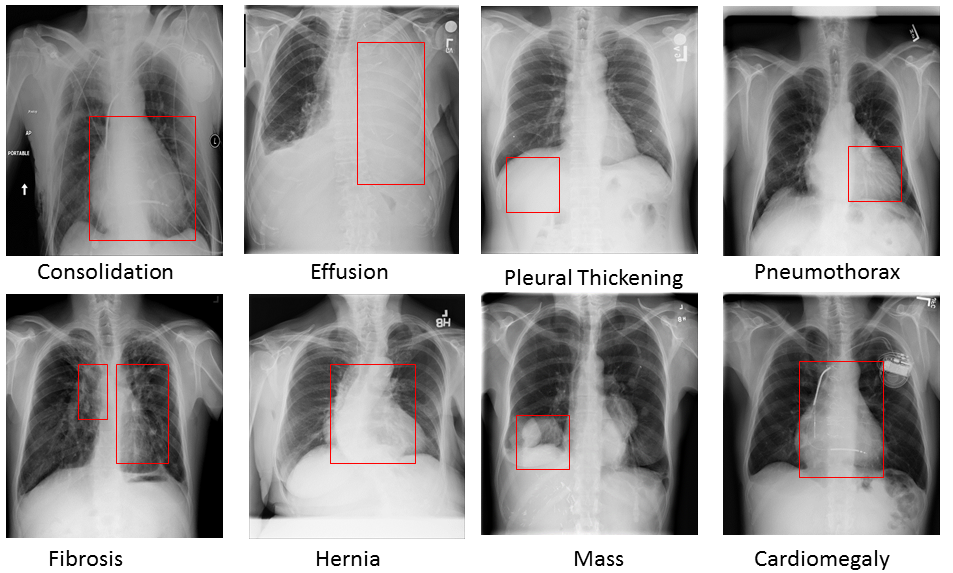}
\caption{Different types of chest abnormalities in CXRs.}
\label{fig:fig1}
\end{figure}

\noindent Lungs are the fundamental organs within the human respiratory system which provide respiration. Lungs are enclosed within the human chest, and their pathology is majorly observed through chest radiographs, also known as chest X-rays (CXRs) \cite{Rabia2018ICIAR}. Apart from this, many pulmonary disorders are also diagnosed through CXRs by observing the abnormal pathological patterns within the poster-anterior, anteroposterior, and lateral projections of the thoracic cavity \cite{WEBSITE:2}. Moreover, CXR imagery is also an effective and low-cost modality for detecting edema, tuberculosis, single or multiple nodules, and pneumonia \cite{3} (as shown in Figure \ref{fig:fig1}). Among these pathologies, the fatal one is pneumonia (especially COVID-19 pneumonia \cite{covid1}), which is clinically identified by observing airspace opacities, lobar consolidation, and interstitial opacities \cite{covid1}.
On the other hand, edema is identified through cephalization of the pulmonary vessels, septal lines, patchy shadowing (with air bronchograms), and increased cardiac size \cite{WEBSITE:6}. Tuberculosis is identified (from CXRs) by observing the consolidations and cavities that are often seen in the upper lung zones (with or without mediastinal or hilar lymphadenopathy) \cite{WEBSITE:4}. In contrast, the nodules appear as a spot within the lung zones as observed in the CXRs \cite{WEBSITE:5}.
\noindent Many researchers have proposed autonomous frameworks to screen pulmonary diseases using CXRs \cite{survey}. The initial methods employed machine learning to recognize different chest abnormalities at the inference stage \cite{CADdate}. However, these methods were confined to limited \TT{datasets} and experimental settings due to the subjectiveness in \TT{their} handcrafted features \cite{5}. With the advances in deep learning, the recent wave of diagnostic frameworks utilizes convolutional neural networks to screen and grade different chest abnormalities like pneumonia \cite{CheXnet}. Deep learning, although, increased the diagnostic performance of such frameworks by many folds. Still, they require exhaustive training efforts involving high computational power and large-scale (and well-annotated) data, and this limits their applicability towards screening new types of pathologies in a clinical setting. The incremental learning paradigm addresses this inherent limitation of deep neural networks. But incremental learning systems are vulnerable to catastrophic forgetting, which is defined as the inability of the classification model to forget its prior knowledge upon learning new tasks \cite{learnwithoutforg, hassan2020Sensors}. To overcome catastrophic forgetting within an incremental learning framework, many researchers have proposed knowledge distillation \cite{ICARL} and contrastive learning \cite{tian} based strategies. But these schemes ignore the structural similarities and interdependencies between different knowledge representations, which can significantly boost the classification performance of the incremental learning systems while showcasing high resistance to catastrophic forgetting. 

\section{Related Work} \label{sec:related}
\noindent Deep learning has increased the performance of the medical image diagnostic frameworks \cite{survey2, LACNN} by many folds, especially to predict abnormal lung pathologies from the CXRs \cite{CheXnet}. The majority of these systems are based on transfer learning or fine-tuning approaches, which utilize pre-trained models such as VGG-19 \cite{vgg}, MobileNet \cite{mobilenet} to classify abnormal chest pathologies \cite{transfer, Hassan2019Sensors, transfer2} using CXRs. Although transfer learning systems can fulfill the large-scale data requirement (to some extent) for recognizing a limited set of chest pathologies \cite{survey, Rabia2018ICIAR, covid1}. However, these systems tend to forget their source domain knowledge (while they are tuned for the target domain tasks) \cite{Rosenfeld2020TPAMI}. Due to this, they cannot be deployed in hospitals for screening purposes. In clinical practice, the classification model is expected to recognize new types of pathologies (while retaining its previously learned knowledge) using very few training examples \cite{ragfw, cbm}. Incremental learning enables the deep learning models to meet this requirement. More specifically, the concept of incremental learning originates to periodically learn scarce yet inter-related tasks effectively without re-training the model from scratch \cite{dfincre}. Furthermore, by showing high resistance to catastrophic forgetting, the deep incremental learning models can eliminate the need for re-training on the large datasets to learn the limited number of classification categories \cite{lesfo, CF}. 

%HERE
\noindent The initial strategies to address catastrophic forgetting phenomena were based on distillation \cite{kd}, in which the knowledge of the previously trained instance of the model is compressed and transferred to the new instance. The new instance then retains these representations by minimizing the distillation loss function explicitly (through the training examples which correspond to the previously added classes) \cite{kd}. Apart from this, the new model instance is also penalized (during training) to learn the new classes (from the provided small-scale set of training examples) \cite{dmc}. Here, the work of Li et al. \cite{learnwithoutforg} is notable where they proposed the learning without forgetting (LwF) scheme. LwF \cite{learnwithoutforg} optimizes the distillation and cross-entropy loss functions in each training increment to enable the model to retain its prior knowledge while learning new representation simultaneously.
Moreover, Rebuffi et al. \cite{ICARL} improved the LwF \cite{learnwithoutforg} for the class-incremental learning tasks by indefinitely learning the feature representations related to the distilled and newly added categories via joint distillation and classification loss function optimization. Aljundi et al.  \cite{expertgate} introduced gating auto-encoders that can incrementally learn the feature representations for the task at hand, and based upon the nature of the test \TT{sample,} the processing request is automatically forwarded to the relevant gate to perform the appropriate classification task. Castro et al. \cite{enetoend} presented an end-to-end incremental learning scheme in which they proposed a cross-distilled loss function to retain prior learned knowledge while learning new classification tasks by penalizing the candidate network via cross-entropy loss function. Similarly, Roy et al. \cite{treecnn} presented hierarchically-fashioned CNN architecture that is incrementally trained to perform various classification tasks with minimal training efforts. Tian et al. \cite{tian} presented a contrastive learning strategy that outperformed knowledge distillation and other cutting-edge distillers for various knowledge transfer tasks, including single model compression, ensemble distillation, and cross-modal transfer. They proposed an objective function that minimizes the Kullback–Leibler (KL) divergence between a teacher and student network's outputs.  
Mirzadeh et al. \cite{Mirzadeh} addressed the gap in teacher-student learning network through a multi-step knowledge distillation process.
Lee et al. \cite{lee} introduced a global distillation strategy to reduce catastrophic forgetting on the massive unlabeled data. Lopez-Paz et al. \cite{GEM} proposed utilizing episodic memories in their framework, dubbed Gradient Episodic Memory (GEM), to resist catastrophic forgetting during incremental training. Chaudhry et al. \cite{agem} proposed averaged GEM (A-GEM) that replaces the quadratic programming (in GEM \cite{GEM}) with dot products through which the gradient calculation is performed only once. \TT{This modification} significantly \TT{reduces} the computational \TT{resources which were required} in the original GEM \cite{GEM} framework. Apart from this, Hegde et al. \cite{Hegde} presented a compact and sparse student network that replicates the compressed representation of the teacher model to learn various classification via knowledge distillation incrementally. 

\noindent Although, the state-of-the-art incremental learning strategies give decent classification performance while overcoming catastrophic forgetting phenomena. Still, these schemes treat the previously learned representations and newly stacked classes independent of each other, and this cap the performance of the deep classification networks towards performing these tasks simultaneously, especially when they are highly correlated and inter-related with each other \cite{tian}. To address these limitations, we present a novel incremental learning loss function ($L_{IL}$) that not only penalizes the classification network to learn newly added class representations while distilling the previously learned knowledge. But it also ensures that the network understands their complex relationships and structural dependencies to recognize them effectively at the inference stage. To summarize, the main contributions of the paper are:

\begin{itemize}
\item This paper presents a first attempt towards utilizing incremental learning to periodically screen different pulmonary disorders from the CXRs irrespective of their scanner specifications.

\item Unlike state-of-the-art incremental learning approaches \cite{ICARL, kd}, the proposed framework analyzes the structural and semantic relationships between periodically learned chest abnormalities (via proposed $L_{IL}$ loss function), which enables it to recognize them effectively at the inference stage.

\item The proposed framework is rigorously evaluated on five public datasets, where it achieved the accuracy and F1 score of up to 0.8405 and 0.8303, respectively. Furthermore, it outperforms the state-of-the-art schemes on all five datasets, as evident from the Section \ref{sec:results}.

\end{itemize}

\noindent The rest of the sections are organized as follows: Section \ref{sec:method} presents the proposed framework in detail. Section \ref{sec:exp} enlists experimental protocols. Section \ref{sec:results} presents the results of the proposed framework and its comparison with the state-of-the-art schemes. Section \ref{sec:conclusion} concludes the paper and envisage future directions.
 
\section{Methodology} \label{sec:method}
\begin{figure}[t]
\includegraphics[width=1\linewidth]{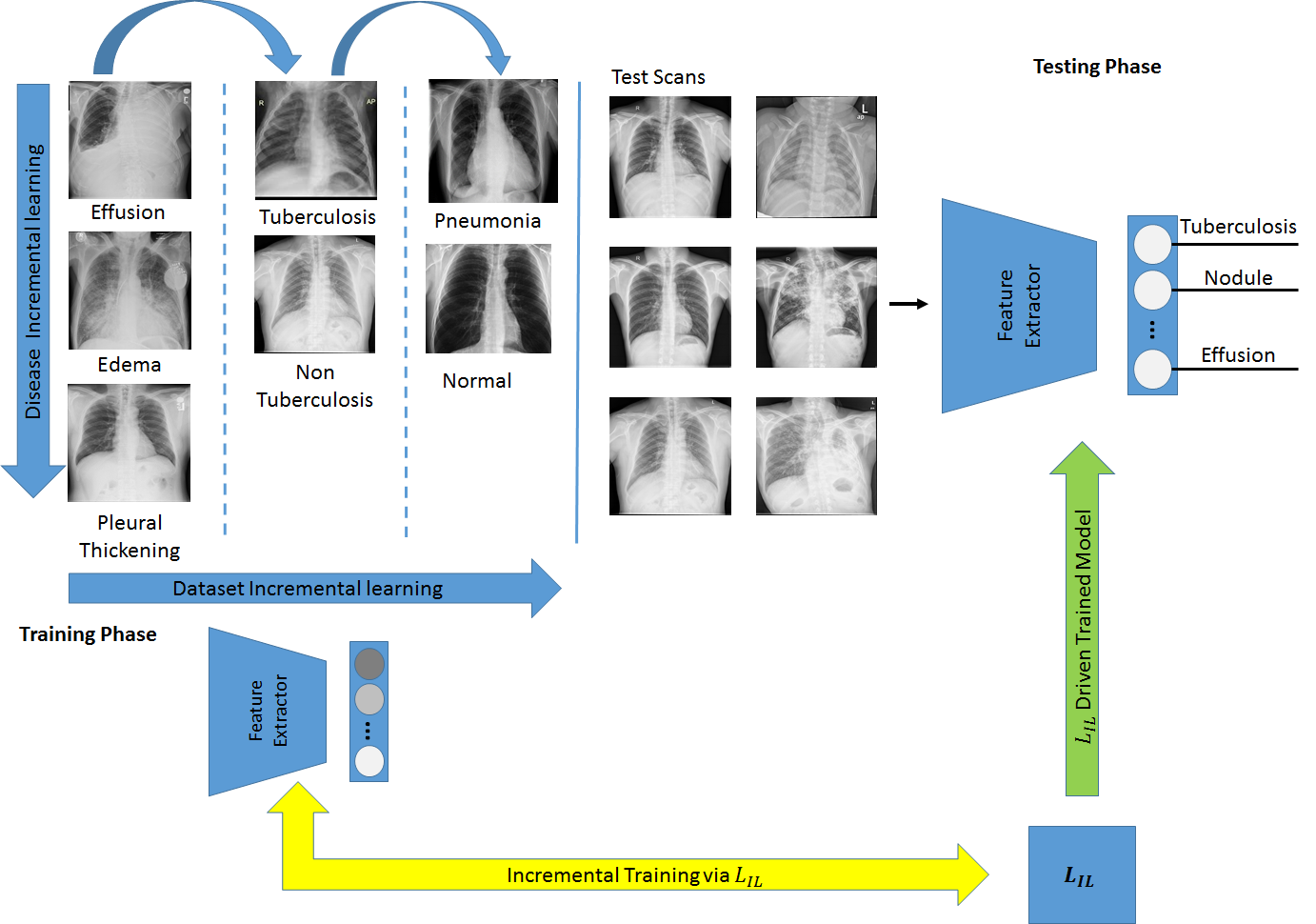}
\caption{Block diagram of the proposed framework. The classification network (within the proposed system) is trained incrementally in two phases. The first phase recognizes different chest abnormalities during incremental training by minimizing the $L_{IL}$ loss function. In the second phase, the classification model learns different pulmonary disorders from multiple datasets (acquired through different scanners). At the inference (testing) stage, the incrementally trained classification network can recognize different chest diseases from the provided test CXR scans simultaneously, irrespective of their scanner specifications.}
\label{fig:fig2}
\end{figure}

\noindent The block diagram of the proposed framework is shown in Figure \ref{fig:fig2}. Here, we can see that the proposed framework is trained in two phases. In the first phase, the classification model (within the proposed framework) is penalized via $L_{IL}$ to recognize various chest abnormalities periodically. We dubbed this phase as disease incremental learning. Moreover, in the second phase, the proposed framework performs dataset incremental learning to train the candidate network (via $L_{IL}$) for recognizing chest diseases from different datasets. At the inference stage, the proposed framework can identify different pulmonary disorders via CXR imagery regardless of their scanner specifications. More details about training and testing phases are presented in the subsequent sections.

\subsection{Training Phase}
\noindent The incremental training of the proposed framework is done in two phases. The first phase is related to the disease incremental learning, and the second phase is related to the dataset incremental learning. 

\subsubsection{Disease Incremental Learning}
\noindent In the first training phase (\TT{related} to the disease incremental learning), we train the candidate classification model (incrementally) to recognize different chest abnormalities such as effusion, pulmonary edema, pleural thickening, etc., from the first dataset. Here, in each training increment, the $L_{IL}$ loss function ensures that the network minimizes the classification loss while learning newly added disease categories. Furthermore, $L_{IL}$ also penalizes the classification network to distill its prior learned knowledge representations through a subset of training examples that were used in the previous iteration. Moreover, unlike other incremental learning schemes \cite{tian, ICARL, learnwithoutforg, kd}, the proposed $L_{IL}$ also minimizes the mutual distillation objective function $L_{MD}$. $L_{MD}$ ensures that the classification model resolves the complex inter-dependencies between incrementally learned knowledge representations via Bayesian inference. More details about the loss functions are presented in Section \ref{sec:loss}.

\subsubsection{Dataset Incremental Learning}
\noindent In the second training phase, the proposed framework is trained to recognize pulmonary disorders across different datasets. Although the training strategy remains the same, i.e., the classification model within the proposed framework is penalized in each training increment (via $L_{IL}$) to recognize different disease categories \TT{from} the CXR scans. However, these CXR scans are acquired with different scanners. Also, these CXR scans belong to different datasets, showcasing different types of abnormal disease patterns. The proposed framework learns such kind of diversified classification tasks (incrementally) without catastrophically forgetting its prior knowledge by minimizing the $L_{MD}$ objective function within $L_{IL}$. Due to $L_{MD}$, the $L_{IL}$ loss function optimizes the classification model, which yields better performance than state-of-the-art incremental learning schemes. The detailed comparison of the proposed framework with existing approaches is presented in Section \ref{sec:results}.

\subsection{Testing Phase}
\noindent At the testing (inference) stage, the proposed framework possesses the capacity to recognize different pulmonary diseases from the CXR scans regardless of their scanner specifications. Unlike the conventional transfer learning or fine-tuning approaches, the proposed framework can be further modified to recognize more chest pathologies (via few-shot training) using the proposed $L_{IL}$ loss function. This makes the proposed framework scalable and an ideal choice for clinical screening, as it can be easily adapted to recognize emerging diseases using very few annotated training examples. 

\subsection{Proposed Incremental Learning Loss Function} \label{sec:loss}
\noindent We propose a novel incremental learning loss function, dubbed $L_{IL}$, which penalizes the candidate classification model to learn new disease classification tasks while simultaneously retaining its prior learned knowledge. Furthermore, the proposed $L_{IL}$ loss function enables the classification model to showcase high resistance to catastrophic forgetting phenomena compared to the recently introduced incremental learning approaches. This is because $L_{IL}$ considers the knowledge representations which the classification models learn to be non-mutually exclusive. Through the proposed $L_{MD}$ objective function, the $L_{IL}$ penalizes the candidate network to exploit the structural and semantic similarities between different incrementally learned knowledge representations, which result in the superior classification performance as compared to the state-of-the-art approaches. Mathematically, the $L_{IL}$ loss function is expressed below: 

\begin{equation}
L_{IL} = \alpha  L_{C}+ \beta L_{MD}+ \gamma L_{D}\TT{,}
\label{eq:eq1}
\end{equation}

\begin{equation}
L_{D}= -\frac{1}{b_{s}}\sum_{k=0}^{b_{s}-1}\sum_{m=0}^{n_o-1}t_{o}^{\tau}(x_{k,m})\log \left(p(l_o^{\tau}(x_{k,m})) \right)\TT{,}
\label{eq:eq2}
\end{equation}
and

\begin{equation}
L_{C}= \frac{1}{b_{s}}\sum_{k=0}^{b_{s}-1}\sum_{m=0}^{n_n-1}q(t_{n}^{\tau}(y_{k,m})) \log \left( \frac{q_{n}^{\tau}(y_{k,m})}{p(l_{n}^{\tau}(y_{k,m}))} \right)\TT{,}
\label{eq:eq3}
\end{equation}

\noindent where $b_{s}$ denotes the batch size, $n_o$ and $n_n$ denotes, respectively, the number of training examples associated with previously learned ($o$) and newly added ($n$) disease categories (in the current training iteration), and $\tau$ is the temperature constant that scales the ground truth labels and the output logits. Apart from this, $t_{o}^{\tau}$ in Eq. \ref{eq:eq2} represents the scaled ground truth labels for the training examples ($x$) belonging to the previously learned categories $o$, and $p(l_o^{\tau}(x_{k,m}))$ is the predicted softmax probability obtained from the scaled logits $l_o$ of the training examples ($x$) belonging to the previously learned categories. In Eq. \ref{eq:eq3}, $q(t_{n}^{\tau}(y_{k,m}))$ denotes the true softmax distribution of the scaled ground truth labels ($t_n^{\tau}$) associated with the training examples ($y$) which enables the classification network to learn newly added classes $n$, and $p(l_{n}^{\tau}(y_{k,m}))$ represents the predicted softmax probability of the scaled output logits $l_n$ generated from training examples $y$ associated with the newly added disease categories (in current training iteration).

\noindent From Eq. \ref{eq:eq2} and \ref{eq:eq3}, we can see that $L_{D}$ is a distillation loss function which ensures that the network does not forget previously learned classes in the current training increment. Moreover, $L_{C}$ is a classification loss function that penalizes the candidate model to learn newly added disease categories through their respective training examples. Moreover, the parameters $\alpha$, $\beta$, and $\gamma$ represent the loss weights which are empirically determined to be 0.5, 0.25, and 0.25, respectively, across all the datasets. Contrary to the existing knowledge distillation approaches, we introduced a novel mutual distillation objective function ($L_{MD}$) in the proposed $L_{IL}$ function. $L_{MD}$ bridges the gap between newly added and previously learned knowledge representations by exploiting their complex structural and semantic dependencies. 

\subsubsection{Mutual Distillation Objective Function}
\noindent Incrementally learned knowledge representations are generally non-mutually exclusive in nature \cite{ILSurvey}. Therefore, penalizing the classification networks to recognize their complex relationships and dependencies is crucial for achieving robust performance and high resistance to catastrophic forgetting during incremental training. To handle this, we introduce a novel $L_{MD}$ function that utilizes Bayesian inference to analyze the extent of similarities between incrementally learned knowledge representations and minimize them to recognize inter-related disease categories effectively at the inference stage. For any classification network having an input $z$ during incremental training, such that $z=\{x, y | x, y \in \mathbb{R}^2 \}$, where $x$ and $y$ denotes the training examples of the previously learned and newly added disease categories, respectively. To analyze the extent of similarities, we compute the joint probability distribution between the scaled output logits $l_o^{\tau}(x)$ and $l_n^{\tau}(y)$, such that $p(l_o^{\tau}(x),l_n^{\tau}(y))=p(l_o^{\tau}(x)|l_n^{\tau}(y))\times p(l_n^{\tau}(y))$. Similarly, we have $p(l_n^{\tau}(x),l_o^{\tau}(y))=p(l_n^{\tau}(x)|l_o^{\tau}(y))\times p(l_o^{\tau}(y))$. Adding the notion of disease categories ($d$) to these definitions yield:

\begin{equation}
p(l_o^{\tau}(x),l_n^{\tau}(y)|d=d_i)=\sum_{k=0}^{n_o-1}p(l_o^{\tau}(x),l_n^{\tau}(y)|d=d_i)p(l_n^{\tau}(y)|d=d_i)\TT{,} 
\label{eq:eq4}
\end{equation}

\begin{equation}
p(l_n^{\tau}(y),l_o^{\tau}(x)|d=d_i)=\sum_{k=0}^{n_n-1}p(l_n^{\tau}(y),l_o^{\tau}(x)|d=d_i)p(l_o^{\tau}(x)|d=d_i)\TT{,}
\label{eq:eq5}
\end{equation}

\noindent Afterwards, the posterior for each class $d_i \in d$ is computed through Bayes rule:

\begin{equation}
p(d=d_i|l_o^{\tau}(x),l_n^{\tau}(y))= \frac{p(l_o^{\tau}(x),l_n^{\tau}(y)|d=d_i) p(d=d_i)} {\sum_{k=0}^{n_o-1}p(l_o^{\tau}(x),l_n^{\tau}(y)|d=d_k)}\TT{,}
\label{eq:eq6}
\end{equation}

and

\begin{equation}
p(d=d_i|l_n^{\tau}(y),l_o^{\tau}(x))= \frac{p(l_n^{\tau}(y),l_o^{\tau}(x)|d=d_i) p(d=d_i)} {\sum_{k=0}^{n_n-1}p(l_n^{\tau}(y),l_o^{\tau}(x)|d=d_k)}\TT{,}
\label{eq:eq7}
\end{equation}

\noindent $p(d=d_i)$ denotes the probabilities of disease $d_i$ occurrence. It is calculated as $p(d=d_i)=\frac{\sum_{j=0}^{n_{di}-1}j}{\sum_{k=0}^{n_d-1}k}$, where $n_{di}$ represents the total number of disease $d_i$, and $n_d$ denotes the total number of disease categories $d$. Moreover, in Eq. \ref{eq:eq6} and \ref{eq:eq7}, $p(d=d_i|l_o^{\tau}(x),l_n^{\tau}(y))$ and $p(d=d_i|l_n^{\tau}(y),l_o^{\tau}(x))$ denotes the posterior probabilities, $p(l_o^{\tau}(x),l_n^{\tau}(y)|d=d_i)$ and $p(l_n^{\tau}(y),l_o^{\tau}(x)|d=d_i)$ denotes the likelihood, respectively. Since these likelihoods derive from the numerical representations, therefore, in the proposed framework, we model them through the multivariate Gaussian distribution, as expressed below:

\begin{equation}
    \mathcal{N}(\hat{z}|\mu,\Sigma) =  \frac{1}{\sqrt{2 \pi^\mathcal{D} |\Sigma|}}  \exp{-\frac{1}{2}(\hat{z}-\mu)^T\Sigma^{-1}}(\hat{z}-\mu)\TT{,}
    \label{eq:eq8}
\end{equation}

\begin{equation}
    \mu = \frac{1}{\zeta} \sum_{i=0}^{\zeta-1}\hat{z}_i; \Sigma=\frac{1}{\zeta-1} \sum_{i=0}^{\zeta-1} (\hat{z}_i-\mu)(\hat{z}_i-\mu)^T\TT{,}
    \label{eq:eq9}
\end{equation}

\noindent where $\hat{z}_i \in z$, $\mathcal{D}$ represents the multivariate dimension, $\mu$ and $\Sigma$ denotes the average and covariance of the output logits, respectively. Afterward, $L_{MD}$ is computed as:

\begin{equation}
\begin{split}
L_{MD}= -\frac{1}{b_{s}}\sum_{k=0}^{b_{s}-1}\sum_{m=0}^{n_o-1}t_{o}^{\tau}(x_{k,m})\log p(d=d_i|l_o^{\tau}(x),l_n^{\tau}(y)) \\
 -\frac{1}{b_{s}}\sum_{k=0}^{b_{s}-1}\sum_{m=0}^{n_n-1}t_{n}^{\tau}(y_{k,m})\log p(d=d_i|l_n^{\tau}(x),l_o^{\tau}(y))\TT{,}
\end{split}
\label{eq:eq10}
\end{equation}
where $t_{o}^{\tau}(x)$ and $t_{n}^{\tau}(y)$ denotes the scaled ground truth labels corresponding to the training samples $x$ and $y$ of previously learned and newly added classes, respectively. 

\section{Experimental setup} \label{sec:exp}
\noindent This section presents a detailed discussion on the experimental protocols (including dataset description, training strategies, and evaluation metrics), which we followed for conducting the proposed study.

\subsection{Datasets}
\noindent The proposed scheme is evaluated on five publicly available datasets containing high-resolution chest radiographs. The detailed description of each dataset is presented below:

\subsubsection{Indiana Dataset}
\noindent The first dataset which we used for the evaluation of the proposed \TT{framework} is the Indiana dataset \cite{indiana}. Indiana dataset was collected from various hospitals affiliated with the Indiana University School of Medicine \cite{indiana}. The complete dataset contains 7,470 frontal and lateral CXRs depicting normal and abnormal pathologies such as cardiac hypertrophy, pulmonary edema, opacity, or pleural effusion. 

\subsubsection{Montgomery County Dataset}
\noindent The second dataset \TT{on which we tested the proposed system} is \TT{the} Montgomery County (MC) dataset \cite{mont}. This dataset was collected from the Department of Health and Human Services in partnership with Montgomery County, Maryland in the United States. The group consisted of 138 frontal chest radiographs from the Montgomery County Tuberculosis Screening Program, of which 80 were normal, and 58 had tuberculosis symptoms. Moreover, the scans within the MC dataset have a resolution of 4020 $\times$ 4892 and 4892 $\times$ 4020 pixels.

\subsubsection{Shenzhen Dataset}
\noindent The third dataset on which we evaluated the proposed framework is the Shenzhen dataset \cite{schz}. The dataset was collected in collaboration with Shenzhen People’s Hospital, Guangdong Medical College, Shenzhen, China. It contains 662 CXR scans, from which 326 represent normal pathologies and 336 are tuberculosis affected.

\subsubsection{Japanese Society of Radiological Technology Dataset}
\noindent The fourth dataset on which we evaluated the proposed framework is the Japanese Society of Radiological Technology (JSRT) dataset \cite{jsrt} that contains normal scans as well as scans that are affected with the pulmonary nodule. JSRT \cite{jsrt} contains 247 CXRs, from which 154 contains pulmonary nodules (100 malignant and 54 benign), and 93 scans have no nodules. All CXR scans within JSRT dataset \cite{jsrt} have a resolution of 2048 $\times$ 2048 pixels, while their color depth is 12 bits.

\subsubsection{Zhang CXR Dataset}
\noindent The last dataset on which the proposed framework is evaluated is the Zhang CXR dataset \cite{zhang}. Zhang dataset is originally designed to classify different retinal diseases via optical coherence tomography (OCT) imagery \cite{zhang}. However, it also contains CXRs depicting healthy and pneumonic pathologies. In the Zhang CXR dataset \cite{zhang}, 3,883 training scans depict pneumonic pathologies while 1,349 scans are from healthy subjects. Similarly, the testing set contains 390 pneumonic, and 234 healthy scans \cite{zhang}. Moreover, all of these scans are arranged within the dataset as per their depicted pathologies.  

\subsection{Training and Implementation Details} \label{sec:training}
\noindent The proposed framework is implemented using TensorFlow 1.14 and Keras 2.0.0 with Python 3.7.4 on the Anaconda platform. Some of the utility functions are also implemented using MATLAB R2020a. The proposed framework's training was conducted in two phases where the candidate classification model minimized the $L_{IL}$ loss function in each iteration. The number of epochs (in each training increment) was 20 (and the number of cycles in each epoch varies as per each dataset). Also, during each training increment, we fed the candidate network with around 20\% of the original training data (where 10\% were used for the distillation process and the remaining 10\% were used to learn the newly added classes). Apart from this, we used ADADELTA \cite{Zeiler2012ADADELTA} as an optimizer, and the training was conducted on the machine with a Core i7-9750H@2.6 GHz processor, 32GB DDR4 RAM, and NVIDIA RTX 2080 Max-Q GPU with cuDNN v7.5 and a CUDA Toolkit 10.1.243. 

\subsection{Evaluation Metrics}
\noindent To evaluate the proposed framework and to compare it with the state-of-the-art schemes, we used the standard classification metrics such as accuracy, true positive rate ($TPR$), positive predicted value ($PPV$), and $F1$ scores, as expressed below:

\begin{equation}
    Accuracy = \frac{T_P+T_N}{T_P+T_N+F_P+F_N}\TT{,}
    \label{eq:eq11}
\end{equation}

\begin{equation}
    TPR = \frac{T_P}{T_P+F_N}\TT{,}
    \label{eq:eq12}
\end{equation}

\begin{equation}
    PPV = \frac{T_P}{T_P+F_P}\TT{,}
    \label{eq:eq13}
\end{equation}

\begin{equation}
    F1 = \frac{2 \times TPR \times PPV}{TPR+PPV}\TT{,}
    \label{eq:eq14}
\end{equation}

\noindent where $T_P$, $T_N$, $F_P$, and $F_N$ denotes the true positives, true negatives, false positives, and false negatives, respectively.

\section{Results} \label{sec:results}
\noindent The proposed framework has been thoroughly evaluated on five public CXR datasets with four different classification networks, i.e., VGG-16 \cite{vgg}, ResNet-50 \cite{resnet}, ResNet-101 \cite{resnet} and MobileNet \cite{mobilenet}. Furthermore, these networks have been trained incrementally where, in each iteration, they minimized the proposed $L_{IL}$ loss function to learn \TT{different pulmonary disease classification tasks}. To present the evaluation results of the proposed framework in the best manner, we organized this section as follows: At first, we present detailed ablation studies to determine the hyperparameters of the proposed framework on all five datasets. Afterward, we present a detailed evaluation of the proposed on each dataset.

\subsection{Ablation Study}
\noindent The ablation study for the proposed framework include 1) the determination of optimal temperature constant $\tau$, and 2) to evaluate the capacity of $L_{IL}$ in resisting the catastrophic forgetting with and without the inclusion of $L_{MD}$ objective function across each dataset. 

\subsubsection{Effect of $\tau$}
\noindent The temperature constant $\tau$ generates soft-target probabilities for each class, \TT{enabling} deep neural networks to \TT{accurately} learn the distinct feature representations during the knowledge distillation process. It should be noted that $\tau$ is a dataset-dependent parameter, and its optimal value varies across different datasets. Table \ref{tab:tau} reports the effect of varying $\tau$ in terms of classification error across each dataset. Here, we can see that although the optimal value of $\tau$ varies across each dataset\TT{, it typically} ranges between $2 \leq \tau \leq 2.5$ for each classification network.

\begin{table}
\footnotesize
    \centering
    \caption{Effect of varying temperature constant $\tau$ in terms of classification error across each dataset. Bold indicates the $\tau$ value with optimal classification performance.}
    \begin{tabular}{c c c c c c c}
        \hline\small
         Dataset & Model & $\tau=1.5$  & $\tau=2$ & $\tau=2.5$ & $\tau=3$ & $\tau=3.5$ \\
         \hline
         Indiana \cite{indiana} & MobileNet \cite{mobilenet} & 0.3648 & 0.3217 & \textbf{0.3064} & 0.3979 & 0.4302 \\
         & ResNet\textsubscript{50} \cite{resnet}	& 0.1968 & \textbf{0.1731} & 0.2584 & 0.3168 & 0.3709 \\
         & ResNet\textsubscript{101} \cite{resnet} & 0.1716 & \textbf{0.1595} & 0.2247 & 0.2984 & 0.3474 \\
         & VGG\textsubscript{16} \cite{vgg} & 0.2431 & 0.2052 & \textbf{0.1818} & 0.2603 & 0.3372 \\
        \hline
        MC \cite{mont} & MobileNet \cite{mobilenet} & 0.4410 & \textbf{0.3742} & 0.3931 & 0.4628 & 0.5214 \\
         & ResNet\textsubscript{50} \cite{resnet}	& 0.3201 & \textbf{0.2972} & 0.3163 & 0.3826 & 0.4517 \\
         & ResNet\textsubscript{101} \cite{resnet} & 0.2741 & \textbf{0.2392} & 0.2645 & 0.3392 & 0.4081 \\
         & VGG\textsubscript{16} \cite{vgg} & 0.4312 & 0.3882 & \textbf{0.3479} & 0.3930 & 0.4593 \\
         \hline
        Shenzhen \cite{schz} & MobileNet \cite{mobilenet} & 0.3809 & \textbf{0.3430} & 0.3673 & 0.4035 & 0.4490 \\
         & ResNet\textsubscript{50} \cite{resnet}	& 0.3513 & 0.2678 & \textbf{0.2478} & 0.2950 & 0.3621 \\
         & ResNet\textsubscript{101} \cite{resnet} & 0.3341 & 0.2536 & \textbf{0.2327} & 0.3164 & 0.3972 \\
         & VGG\textsubscript{16} \cite{vgg} & 0.3602 & \textbf{0.2825} & 0.3154 & 0.3747 & 0.4286 \\
         \hline
        JSRT \cite{jsrt} & MobileNet \cite{mobilenet} & 0.5432 & 0.4921 & \textbf{0.4778} & 0.5329 & 0.5832 \\
         & ResNet\textsubscript{50} \cite{resnet}	& 0.3968 & 0.3650 & \textbf{0.3442} & 0.4001 & 0.4527 \\
         & ResNet\textsubscript{101} \cite{resnet} & 0.3427 & 0.3126 & \textbf{0.3037} & 0.3697 & 0.4126 \\
         & VGG\textsubscript{16} \cite{vgg} & 0.4163 & 0.3902 & \textbf{0.3847} & 0.4291 & 0.4523 \\
         \hline
        Zhang \cite{zhang} & MobileNet \cite{mobilenet} & 0.2915 & \textbf{0.2581} & 0.2721 & 0.3283 & 0.3749 \\
         & ResNet\textsubscript{50} \cite{resnet}	& 0.2539 & \textbf{0.1795} & 0.2014 & 0.2493 & 0.3109 \\
         & ResNet\textsubscript{101} \cite{resnet} & 0.2261 & \textbf{0.1475} & 0.1985 & 0.2302 & 0.2904 \\
         & VGG\textsubscript{16} \cite{vgg} & 0.2903 & 0.2432 & \textbf{0.2276} & 0.2846 & 0.3386 \\
         \hline
    \end{tabular}
    \label{tab:tau}
\end{table}

\subsubsection{Effect of $L_{MD}$ in $L_{IL}$ Loss Function}
\noindent The second ablation study analyzes the effect of $L_{MD}$ within the proposed $L_{IL}$ loss function. From Table \ref{tab:lmd}, we can observe that for all datasets, including $L_{MD}$ objective function within $L_{IL}$ significantly improved the classification performance of the proposed framework in terms of accuracy. This is because of the fact $L_{MD}$ enables $L_{IL}$ to analyze the complex inter-dependencies between different incrementally learned knowledge representations, which makes the candidate classification model to effectively screen different inter-related disease categories while showcasing high resistance to catastrophic forgetting.  

\begin{table}
\footnotesize
    \centering
    \caption{Effect of including $L_{MD}$ within the proposed $L_{IL}$ loss function in terms of classification accuracy.}
    \begin{tabular}{c c c c c c }
        \hline\small
         Dataset & Metric & MobileNet \cite{mobilenet} & ResNet\textsubscript{50} \cite{resnet} & ResNet\textsubscript{101} \cite{resnet} & VGG\textsubscript{16} \cite{vgg} \\
         \hline
         Indiana \cite{indiana} & $L_{IL}$ with $L_{MD}$ & 0.6936	& 0.8269 & 0.8405 & 0.8182 \\
         & $L_{IL}$ without $L_{MD}$	& 0.6234	& 0.7624 & 0.7935 & 0.7514 \\
        \hline
         MC \cite{mont} & $L_{IL}$ with $L_{MD}$ & 0.6258	& 0.7028 & 0.7608 & 0.6521\\
         & $L_{IL}$ without $L_{MD}$	& 0.5731	& 0.6249 & 0.6854 & 0.5937 \\
         \hline
         Shenzhen \cite{schz} & $L_{IL}$ with $L_{MD}$ & 0.6570	& 0.7522 & 0.7673 & 0.7175 \\
         & $L_{IL}$ without $L_{MD}$	& 0.5829	& 0.6738 & 0.6903 & 0.6392\\
         \hline
         JSRT \cite{jsrt} & $L_{IL}$ with $L_{MD}$ & 0.5222	& 0.6558 & 0.6963 & 0.6153\\
         & $L_{IL}$ without $L_{MD}$ & 0.4425	& 0.5493 & 0.6068 & 0.5280 \\
         \hline
         Zhang \cite{zhang} & $L_{IL}$ with $L_{MD}$  & 0.7419 & 0.8205 &  0.8525  & 0.7724  \\
         & $L_{IL}$ without $L_{MD}$ & 0.6538	& 0.7112 & 0.7603 &   0.6844 \\
        \hline
    \end{tabular}
    \label{tab:lmd}
\end{table}

\subsection{Evaluation on Indiana Dataset}
\noindent The first dataset for which we evaluated the proposed loss function is the Indiana dataset \cite{indiana} that contains CXR scans depicting cardiac hypertrophy, pulmonary edema, opacity, and pleural effusion. The classification performance of the $L_{IL}$ driven classification models is reported in Table \ref{tab:eval} where we can see that the best performance is achieved for ResNet\textsubscript{101} \cite{resnet}. Also, it should be noted that the performance of the incremental ResNet\textsubscript{101} model is competitive with its fine-tuning variant, i.e., the performance of incremental ResNet\textsubscript{101} model only lags by 10.60\% in terms of accuracy with the fine-tuning approach. 

\begin{table}
\footnotesize
    \centering
    \caption{Evaluation of the proposed framework (with different classification networks). It should be noted here that the results of the fine-tuning approach are achieved using ResNet\textsubscript{101} \cite{resnet} (trained with categorical cross-entropy loss function).}
    \begin{tabular}{c c c c c c c}
        \hline\small
         Dataset & Metric &  MobileNet \cite{mobilenet} & ResNet\textsubscript{50} \cite{resnet} & ResNet\textsubscript{101} \cite{resnet} & VGG\textsubscript{16} \cite{vgg} & Fine-Tuning \\
         \hline
         Indiana \cite{indiana} & Accuracy & 0.6936	& 0.8269& 0.8405& 0.8182& 0.9153\\
         & TPR	&0.6159	&0.8259& 0.8526& 0.7789& 0.9536 \\
         & PPV &	0.6072	&0.7951	&0.8092	&0.7829&	0.8956 \\
         & F1&	0.6269&	0.8102&	0.8303&	0.7808&	0.9236 \\
        \hline
         MC \cite{mont} & Accuracy & 0.6258	& 0.7028& 0.7608& 0.6521& 0.8840\\
         & TPR	&0.6034	&0.6379& 0.7068& 0.5689& 0.8965 \\
         & PPV &	0.5555	&0.6491	&0.7192	&0.5892& 0.8387 \\
         & F1&	0.5784 &	0.6434&	0.7129&	0.5788&	0.8666 \\
         \hline
         Shenzhen \cite{schz} & Accuracy & 0.6570	& 0.7522& 0.7673& 0.7175& 0.8111\\
         & TPR	&0.6398	&0.7232& 0.7440& 0.6994& 0.7857 \\
         & PPV &	0.6697	&0.7783	&0.7861	&0.7320& 0.8328 \\
         & F1&	0.6544 &	0.7497&	0.7644&	0.7153&	0.8085 \\
         \hline
         JSRT \cite{jsrt} & Accuracy & 0.5222	& 0.6558& 0.6963& 0.6153& 0.7894\\
         & TPR	&0.5129	&0.6493& 0.6818& 0.6038& 0.7987 \\
         & PPV &	0.6475	&0.7633	&0.8015	&0.7322& 0.8541 \\
         & F1&	0.5723 &	0.7016&	0.7368&	0.6618&	0.8254 \\
         \hline
         Zhang \cite{zhang} & Accuracy  & 0.7419  & 0.8205 &  0.8525  & 0.7724  &  0.9118 \\
         & TPR &  0.7794	& 0.8230 & 0.8435&   0.7974 &0.9076    \\
         & PPV &	0.8021 &  0.8819 & 0.9138 & 0.8315 & 0.9490     \\
         & F1   &0.7905	&0.8514	&0.8772	 &0.8140 &0.9278	 \\
        \hline
    \end{tabular}
    \label{tab:eval}
\end{table}

\subsection{Evaluation on MC Dataset}
\noindent After evaluating the proposed framework on Indiana dataset \cite{indiana}, we trained it incrementally on the MC dataset \cite{mont} for screening Tuberculosis subjects. The performance of the proposed framework after adapting to the MC dataset is shown in Table \ref{tab:eval}. Here, we can observe that the classification performance for all the backbone models is similar, except for the ResNet\textsubscript{101} model, which is lagging from its fine-tuning variant by 13.93\%.  

\subsection{Evaluation on Shenzhen Dataset}
\noindent The third dataset on which we evaluated the proposed framework is the Shenzhen dataset \cite{schz}. It can be observed from Table \ref{tab:eval} that on Shenzhen dataset \cite{schz}, all the pre-trained models have similar classification performance (except for the MobileNet \cite{mobilenet}) in terms of accuracy. Also, the best performing incremental ResNet\textsubscript{101} \cite{resnet} model only lags from its fine-tuned variants by 5.40\% which is appreciable. 

\subsection{Evaluation on JSRT Dataset}
\noindent The next dataset on which we evaluated the proposed framework is the JSRT \cite{jsrt}. In Table \ref{tab:eval}, we report the classification performance of the pre-trained models towards recognizing a diverse range of pulmonary disorders. We can observe that the incrementally trained ResNet\textsubscript{101} \cite{resnet} model was able to cope up with the conventional fine-tuning approach by 88.20\% (\TT{i.e.,} it is only lagging from fine-tuned baseline by 11.79\%). Also, it should be noted that using the proposed $L_{IL}$ loss \TT{function,} we incrementally trained our model with very few training examples (as described in Section \ref{sec:training}). Following the same training dataset quota for fine-tuning approach would result in a drastic decrease in performance (for all classification networks) due to over-fitting. Therefore, considering the fact that we achieved good generalization over a diverse range of chest pathologies and scanner specifications with few-shot training, we believe that the performance of $L_{IL}$ driven incremental classification models is significant.

\subsection{Evaluation on Zhang Dataset}
\noindent The last dataset on which we evaluated the $L_{IL}$ driven classification models is the Zhang CXR dataset \cite{zhang}. Unlike other datasets, the Zhang CXR dataset \cite{zhang} can only be utilized for binary classification tasks, i.e., to classify healthy and pneumonic pathologies. Therefore, to evaluate the capacity of the proposed loss function on the Zhang CXR dataset \cite{zhang}, we first trained all the classification models for recognizing only the healthy pathologies (i.e., we tuned all the models for a one-category classification task). Afterward, we incrementally molded them to recognize pneumonic pathologies via few training examples, where all the models are penalized so that they accurately learn distinct feature representation for the pneumonic pathologies (within CXR scans) while retaining their previous knowledge about the healthy pathologies. The classification performance of all the four classification networks on \TT{the} Zhang CXR dataset \cite{zhang} is presented in Table \ref{tab:eval}. Here, we can see that the $L_{IL}$ driven ResNet-101 \cite{resnet} only lags from its fine-tuned variant by 6.50\% in terms of accuracy and 5.45\% in terms of F1 score. It should also be noted here while the incremental ResNet-101 \cite{resnet} uses a significantly lower amount of training samples compared to its fine-tuning variant, i.e., it just uses 522 (134 healthy and 388 pneumonia affected) scans out of 5,232 to identify these pathologies. 

\subsection{Comparison with Incremental Learning Schemes}
\noindent Apart from evaluating the applicability of the proposed $L_{IL}$ loss function with different classification networks, we also compared it with the state-of-the-art incremental learning schemes (based on knowledge distillation \cite{Mirzadeh, ICARL} and contrastive learning \cite{tian}). The comparison is reported in Table \ref{tab:ilcom}. We can observe here that the incremental ResNet\textsubscript{101} model trained using the proposed $L_{IL}$ loss function outperformed the state-of-the-art schemes for each dataset. Furthermore, we can also appreciate the significance of $L_{MD}$ objective function within $L_{IL}$ through Table \ref{tab:ilcom}, where $L_{MD}$ allowed the candidate classification model to understand the mutual information between incrementally learned disease representations that improved its performance by many folds on each dataset (especially on Shenzhen \cite{schz}, JSRT \cite{jsrt}, and Zhang CXR \cite{zhang}).

\begin{table}
\footnotesize
    \centering
    \caption{Performance comparison of the proposed $L_{IL}$ loss function with state-of-the-art knowledge distillation and contrastive learning approaches in terms of classification accuracy on all five public datasets. For fairness, all the schemes have been tested using ResNet\textsubscript{101} \cite{resnet} model. Moreover, bold indicates the best performance on each dataset, whereas the second-best performance is underlined.}
    \begin{tabular}{c c c c c c}
        \hline\small
         Loss Function &  Indiana \cite{indiana} & MC \cite{mont} & Shenzhen \cite{schz} & JSRT \cite{jsrt} & Zhang \cite{zhang} \\
         \hline
         Proposed with $L_{MD}$ & \textbf{0.8405} & \textbf{0.7608} & \textbf{0.7673} & \textbf{0.6963} & \textbf{0.8525} \\
         Proposed without $L_{MD}$ & \underline{0.7935} & \underline{0.6854} & 0.6903 & 0.6068 & 0.7603 \\
         TAKD \cite{Mirzadeh} & 0.7322 & 0.6044 & 0.6163 & 0.5748 & 0.6762 \\
         CRD \cite{tian} & 0.6706 & 0.5434 & \underline{0.7069} & 0.6153 & \underline{0.7884} \\
         iCaRL \cite{ICARL} & 0.6037 & 0.5072 & 0.6918 & \underline{0.6356} & 0.7243 \\
        \hline
    \end{tabular}
    \label{tab:ilcom}
\end{table}

\subsection{Qualitative Evaluations}
\begin{figure}[t]
\centering
\includegraphics[width=0.85\linewidth]{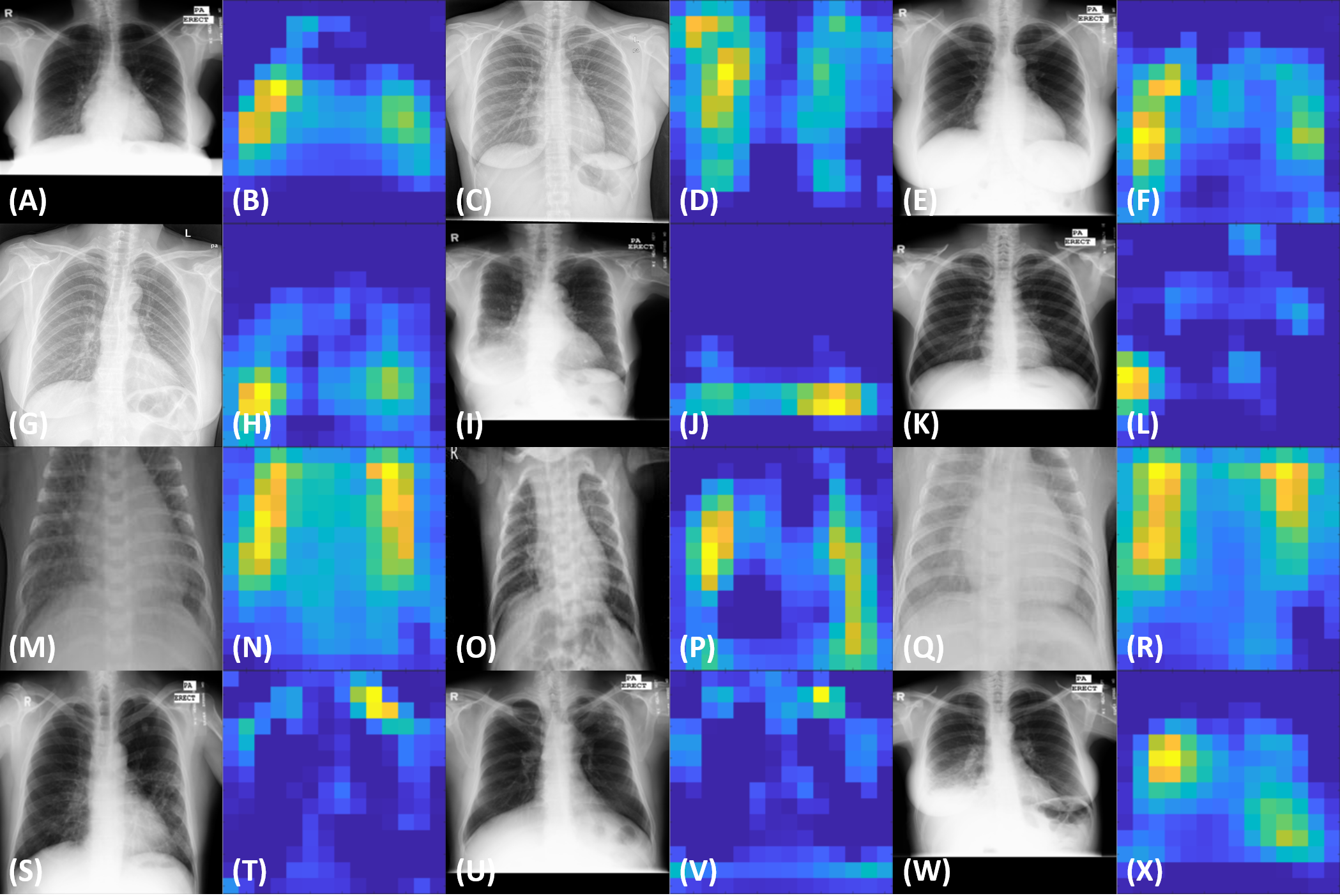}
\caption{Attention maps depicting how the incrementally trained ResNet\textsubscript{101} \cite{resnet} identifies various pathologies across the diverse ranging multi-vendor CXR scanners. Row 1 shows the healthy pathologies, Row 2 shows mixed healthy and nodular pathologies, Row 3 shows pneumonic pathologies, and Row 4 shows tuberculosis and effusion-affected CXRs. The ResNet\textsubscript{101} \cite{resnet} model is trained incrementally using the proposed $L_{IL}$ loss function.}
\label{fig:attention}
\end{figure}

\noindent The qualitative evaluations of the proposed framework is performed by observing the attention maps (obtained by the best performing incremental ResNet\textsubscript{101} \cite{resnet} model) as shown in Figure \ref{fig:attention}. These attention maps are generated from the latent vectors (feature maps within the deeper layers of the network), which are then resized to the network input size. From Figure \ref{fig:attention}, we can observe that the ResNet\textsubscript{101} \cite{resnet} (trained incremental using the proposed $L_{IL}$ loss function) focuses on recognizing the chest abnormalities while predicting the disease categories. For example, see the attention map in Figure \ref{fig:attention} (R), where the network picked the consolidations compared to other scan regions. Similarly, see how the network paid attention to the opacities in Figure \ref{fig:attention} (X). However, not all the focused areas (within attention maps) are clinically relevant. For example, see the focused areas in Figure \ref{fig:attention} (L, T, and V). The incrementally trained network \TT{pays attention to} these irrelevant areas because it could not differentiate between the \TT{lesions} and the background \TT{regions} due to their high spatial similarities \TT{within the candidate scan}. However, it should \TT{also} be noted that although these focused features \TT{are not} clinically relevant, they \TT{do} enable the classification model to predict the disease category (within each scan) \TT{accurately}.

\section{\TT{Conclusion}} \label{sec:conclusion}
\noindent This paper presents a novel incremental learning scheme that can screen various pulmonary diseases from the CXR scans irrespective of their scanner \TT{specifications}. Furthermore, unlike its competitors based on conventional transfer learning and fine-tuning approaches, the proposed framework can effectively recognize new types of chest pathologies with few-shot training without catastrophically forgetting its previously acquired knowledge. The classification network within the proposed scheme is trained via a novel $L_{IL}$ loss function that not only penalizes the network to learn new class representation while distilling its previous knowledge. \TT{Rather, it also ensures that the network understands the complex relationships and inter-dependencies between different knowledge transfer tasks to differentiate them effectively at the inference stage}. Compared to the state-of-the-art incremental learning approaches based on knowledge distillation \cite{Mirzadeh, ICARL}, and contrastive learning \cite{tian}, the proposed framework show high resistance to catastrophic forgetting phenomena. This also results in better classification performance, as evident from Table \ref{tab:ilcom}. Furthermore, the proposed framework has the potential to be deployed in a clinical setting. The computer-aided screening systems (in clinical practice) are expected to recognize new types of pathologies (especially the rarely seen ones) with few training examples. The proposed framework is an ideal choice for such situations, where it can be easily modified to incrementally learn different disease patterns within the CXR scans, unlike the conventional transfer learning approaches. Although, after incrementally learning a diversified range of chest pathologies from five public datasets, the proposed framework showcased some of the clinically irrelevant features for diagnosing the disease categories. For example: see the pairs (K, L), (S, T), (U, V) in Figure \ref{fig:attention}. Nevertheless, the proposed framework does manage to identify these diseases at the inference stage correctly. In the future, we envisage investigating the proposed framework for \TT{screening} COVID-19 and \TT{grading its severity as per the clinical standards}. 

\section*{Acknowledgement}
\noindent We would like to acknowledge the National University of Sciences and Technology, Pakistan, and Khalifa University, UAE, for providing us with the resources in order to conduct this research.

\section*{Conflict of Interest Statement}
\noindent All the authors declare that there are no competing interests that could influence the work presented in this article.

\bibliographystyle{ieeetr}
\bibliography{main}

\end{document}